\documentclass[acmsmall,screen,nonacm]{acmart}

\setcopyright{none}
\renewcommand\footnotetextcopyrightpermission[1]{}

\usepackage{graphicx}
\usepackage{amsmath}

\usepackage{amssymb}
\usepackage{booktabs}
\usepackage{ragged2e}
\graphicspath{{Artworks/}}

\title{Emerging memory technologies at room/cryogenic temperature}

\author{Siddhartha Raman Sundara Raman}
\affiliation{
  \institution{The University of Texas at Austin}
  \city{Austin}
  \state{Texas}
  \country{USA}
}

\begin{abstract}
As conventional technology scaling approaches physical and power limitations, modern computing systems increasingly face performance bottlenecks arising from memory latency, energy consumption, scalability constraints, and data movement overheads. Simultaneously, emerging workloads such as machine learning, graph analytics, and scientific computing demand memory technologies with higher bandwidth, lower latency, improved energy efficiency, and greater storage density. These challenges have motivated extensive research into both room-temperature memories and cryogenic memory systems targeted toward superconducting and quantum computing platforms.

This chapter presents an overview of volatile and non-volatile memory technologies operating across room-temperature and cryogenic environments. The discussion includes SRAM, DRAM, embedded DRAM (eDRAM), NAND/NOR Flash, Resistive Random Access Memory (RRAM), Magneto-resistive Random Access Memory (MRAM), and Ferroelectric Field-Effect Transistor (FeFET)-based memories. In addition, cryogenic technologies including UTBB-SOI-based pseudo-static storage circuits and Josephson Junction Field-Effect Transistor (JJFET)-based devices are discussed in the context of ultra-low-temperature computing systems. The chapter highlights the operational principles, read/write mechanisms, retention behavior, and tradeoffs among area, performance, scalability, and energy efficiency across these memory technologies, while examining challenges and opportunities for future room-temperature and cryogenic computing architectures.

\end{abstract}

\keywords{Memory, SRAM, DRAM, Non-volatile memory, NAND/NOR, RRAM, FeFET, MRAM \\This is an extension of chapter of book published in 2023}

\begin{document}

\maketitle

\section{Introduction} 
\justifying
Modern day systems typically consist of a central processing unit responsible for performing arithmetic and logical operations on the data stored in memory. This architecture is called the Von-Neumann architecture, wherein there are dedicated logic and arithmetic units inside the CPU and the data is read/written from/into memory, and has been the foundational architecture for high performance CPUs till date. A brief background of these CPUs suggest that these processors have undergone tremendous improvements beginning from in-order CPUs that execute the instructions in program order, to modern-day out-of-order CPUs that execute instructions as soon as they are ready to execute and still give an impression to the software that the instructions were executed in-order. This has resulted in an increased performance, with frequencies ranging as high as GHz. However, the major bottleneck from improving the performance further has been the tremendously low memory performance \cite{JJFET} \cite{CIM}\cite{CIM_1}, often referred to as the memory-wall bottleneck \cite{CIM_2}. To alleviate this, an in-depth analysis of existing/emerging memory technologies is required to understand the problems in each of the technologies and further propose solutions. Therefore, in this chapter, we give a brief overview on the existing technologies and the solutions proposed in the literature to understand the tradeoffs made between energy, area and performance. The technologies described are (i) Static Random Access Memory (SRAM), (ii) Dynamic Random Access Memory (DRAM), (iii) NAND/NOR flash (iv) Resistive Random Access Memory (RRAM), (v) Magneto-resistive Random Access Memory (MRAM), (vi) Ferroelectric Field effect Transistor (FeFET). The major reason behind choosing these devices as the case study for non-volatile memories is that there has been a tremendous growth in the recent times about potential of replacing the existing memory topologies with these devices, as they offer bitcell density advantage.

\section{Memory technologies}
The memory technologies can be classified into volatile and non-volatile memory technologies depending on whether the data is retained in memory in the absence of power supply.  From an architectural perspective, processors use caches made of SRAM that are responsible for fast memory accesses and main-memory made of DRAM optimized for density/cost. The hard disk/secondary storage are made of non-volatile memory technologies like NAND Flash and are further optimized for higher density/lower cost. 
\subsection{Volatile Memory technology}
The major memory technologies that are present in the modern-day computers are made of SRAM, DRAM and embedded DRAM (eDRAM). SRAMs have been the workhorse of high performance caches, as they have low access latencies as compared to other memory technologies. Commodity DRAMs have been used in main memory storage, as they have the advantages of high density and a simple bitcell structure. Embedded DRAMs have started to gain traction for caches as they offer higher performance (than commodity DRAM) and high density with a simple bitcell structure, which will be discussed in detail in this section .
\begin{figure}[t]\centering
	\includegraphics[width=\textwidth]{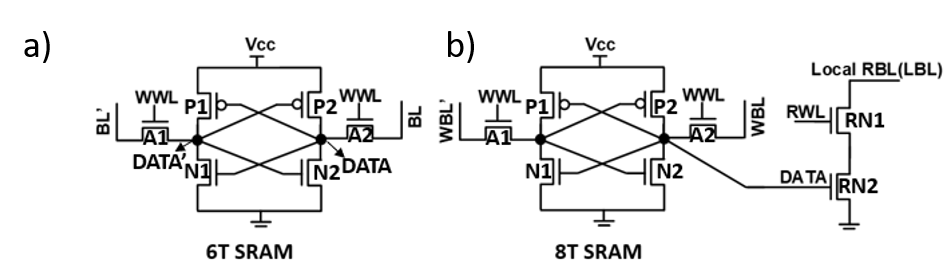}
	\caption{a) 6T SRAM, with a shared read/write port, b) 8T SRAM with decoupled read and write ports} \label{SRAMs}
\end{figure}
\subsubsection{Static Random Access Memory (SRAM)}
SRAMs are classified into 6T, 7T, 8T,9T and 10T structures\cite{SRAM_overall}. The most commonly used SRAM bitcells are the 6T and 8T bitcells. Both these bitcells make use of a cross-coupled inverter as the storage element. The 6T structure is used when overall SRAM area is constrained with high-performance requirements. However, the 8T structure is used in high-performance cache designs, wherein there are no density constraints, because they have the advantage of performing read after write in back-to-back cycles\cite{SRAM_cache}. This section elaborates the tradeoffs in both these designs with respect to performance, area and energy.

\par{\textbf{6T SRAM}}: This design makes use of a shared read-write port, thereby ensuring that either read/write can be performed in a cycle, therefore making it a 1R(read)W(write) port design. 6T SRAMs shown in Fig.\ref{SRAMs} consists of a cross-coupled inverter with pull-up PMOS transistors named P1, P2 and pull-down NMOS transistors named N1, N2 and the NMOS access transistors named A1, A2. In the case of a write operation, data is written onto the bitcell by conditioning the bit lines (BL) to the data value that needs to be stored in the bitcell \cite{SRAM_intro}\cite{NEMGNN_arxiv}\cite{SPARK_ILP}. For instance, for writing '1' onto the bitcell, BL is driven high with a voltage of Vcc and BL\textsuperscript{'} is driven low with a voltage of 0. WWL voltage is driven high so that the DATA node holds a value of '1' and DATA' node holds a '0'. This operation can be split further into 2 phases, namely the initiation and completion phase. The initiation of writing '1' onto the DATA node begins by writing '0' onto the DATA' node, as the NMOS access transistor A1 can pass a good '0', as opposed to NMOS access transistor A2 passing a good '1'. Thus, DATA' going low would imply that the PMOS P2 is slowly being turned ON, thus helping A2 to further write a good '1' onto the bitcell, even though A2 does not pass a good '1' during the completion phase. In order to accomplish a good initiation, the access transistor marked A1 should have a high drive strength as compared to PMOS P1 so that BL' is written successfully onto the DATA' node. In the case of writing a '0' onto DATA node, A2 initiates the process and is completed by the PMOS P1 driving DATA node to Vcc\cite{SRAM_2}. In this case, the design constraint is that the drive strength of A2 must be greater than the drive strength of P2 to accomplish a successful completion. Generalizing this, the initiation is brought about by the node storing '1' and completed by the node storing '0' in the case of a write operation. 
\par The read is preceded by a precharge operation, wherein the BL is precharged to Vcc. During the read operation, with '0' being stored in the bitcell, the BL discharges through A2, with the WWL turned ON and the difference in voltage between BL and BL' is measured using a sense amplifier, that is sensitive to voltage differences as low as 100mV\cite{SA_offset}\cite{SRAM_1} in the modern-day SRAM design. The sense amplifier is typically realized by another cross-coupled inverter design, that is designed to offer precise outputs even in the presence of process variations, often leading to design complexity/overhead. The design constraint during the read operation is that the NMOS transistor should be strong enough to hold the DATA node at '0' even though BL attempts to write a '1' through the access transistor A2.
\par During the retention phase, the WWLs and BLs are turned OFF and data leaks from the node storing '1'\cite{SRAM_1}. This occurs because of the voltage difference between BL and DATA node, with the possibility of a bit-flip, making SRAMs volatile. One common approach is to drive the WL to slightly negative voltages to reduce the leakage through the access transistor. Furthermore, the bitcell Vcc cannot be reduced to 0 even during retention, as this would lead to corruption of bitcell contents. Therefore, the bitcell Vcc needs to be maintained at a minimum voltage, leading to an increased power consumption of the processor \cite{SRAM_1}. This voltage limits the minimum operating voltage of the overall processor design and is often referred to as Vmin.  The issues with 6T SRAM design can be summarized as complex design constraints required for read, write, retention operation in 6T SRAMs, the need for a separate precharge cycle before the read operation.   
\par{\textbf{8T SRAM}}: In an attempt to relax the design constraints for read operation and to amortize the cost of a precharge cycle by performing another "useful" task simultaneously, 8T SRAMs were proposed \cite{SRAM_2}. These have decoupled read/write ports, wherein the access transistors A1 and A2 are used during the write operation and the transistors RN1 and RN2 are used during the read operation, as shown in Fig.\ref{SRAMs}. The major advantage of 8T SRAM is that precharge of RBL attached to the read port can be done in parallel to the write operation done using the access transistors A1, A2. This can be extremely useful in caches, wherein a commonly seen operation is a read followed by write, for which precharge can be overlapped with write, thereby making the read-after-write a 2 cycle operation, as opposed to it being a 3 cycle operation. Furthermore, because the read port transistors are decoupled from write port transistors, the read is independent of the strength of the N2 and A2 transistors, unlike the 6T SRAM scenario, thereby enabling better design characteristics. The write/retention operation in this scenario is the same as that of 6T SRAM\cite{SRAM_8T_2}\cite{SACHI}\cite{NEM_GNN}\cite{SPARK}\cite{Ising_arxiv}.
\par The read operation progresses as: (1) RBL is precharged, (2) RWL is turned ON (3) RBL discharge is sensed.  Unlike the case of 6T SRAM, 8T SRAM involves full swing discharge of RBL during a read operation, therefore enabling realization of the sense amplifier using digital logic gates like a simple 2-input AND gate, with one input being connected to RBL and the other input connected to a reference voltage driven to Vcc.  Furthermore, this form of sensing is called single ended sensing as RBL' is not utilized in sensing, unlike the case of 6T. A design optimization for improving performance is that instead of sharing the RBL across all the rows in a column, RBL that is sensed for read in the case of 8T SRAM, is shared across only a few rows in the column (indicated as local RBL), although there is still a RBL shared across all rows in the column (called the global RBL). During the precharge operation, "global RBL" is initially precharged to Vcc, which further precharges "local RBL" to Vcc \cite{SRAM_8T_3}. However, during the read operation, local RBL discharges and the global RBL is used as the reference voltage for the sensing operation. This decoupling of global and local RBL helps in reducing the discharge latency/read time improving the performance further, as there is lesser capacitance to discharge as opposed to having a higher capacitance, when discharging a complete column \cite{8T_SRAM}. 
\begin{figure}[t]\centering
	\includegraphics[width=\textwidth]{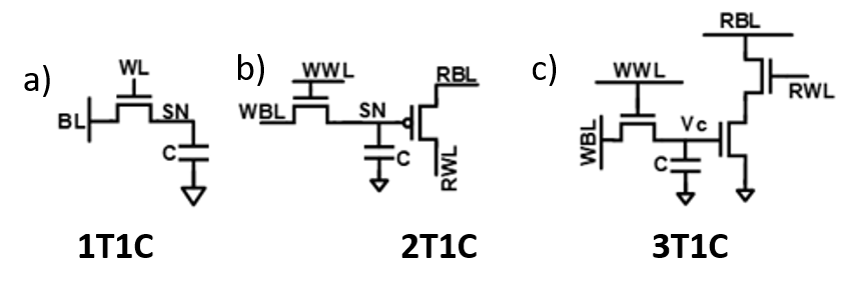}
	\caption{a) DRAM variants without/b),c) with decoupled read and write ports with SN/Vc showing the storage node} \label{DRAMs}
\end{figure}
\subsubsection{Dynamic Random Access Memories (DRAM)}
Commodity DRAMs, used as the main memory storage structure, has the advantages of high storage density owing to its small bitcell size. The bitcell is made of 1T1C (1 Transistor, 1 Capacitor) structure (Fig.\ref{DRAMs})\cite{CDRAM}, with the capacitor used as the storage element. The capacitor is typically a deep trench capacitor, with its capacitance value in the order of 10s of fF, in order to store the charge with minimal leakage. However, it is still a volatile memory because of the discharge from the "leaky capacitor", requiring periodic refreshes to refresh the data that is stored. This bitcell is optimized for cost, but is manufactured separately, integrated with the processor on a separate chip, leading to long latencies for DRAM accesses \cite{DRAM}\cite{PDN}.
\par{\textbf{1T1C}}: The write operation is accomplished by turning on the WL and driving the BL to the necessary voltage value, thereby charging the bitcell capacitor. The read operation is carried out by first precharging the BL to Vcc/2, turning on WL and using a sense amplifier to sense the voltage drop/increase on the BL \cite{CDRAM_2}. It is important to note that the read operation naturally carries out the refresh operation as the sense amplifier is connected to the BL, therefore driving SN to the sensed value. This refresh action is important because the read of the bitcell is disruptive, as the SN voltage changes during the discharge of BL. For instance, during read of '0', SN settles to a value between 0-Vcc/2 and during read of '1', SN settles to a value between Vcc/2-Vcc. The disadvantage of this design is that the discharge of BL is susceptible to process variations and may lead to inaccurate computations, if the data is not restored/refreshed back by the sense amplifier. During the retention phase, WL is turned off/driven to negative voltages to reduce the leakage through the access transistor. 
\par{\textbf{2T1C}}: Although the commodity DRAMs still make use of 1T1C structure, bitcells have been proposed to eliminate the disruptive read. One such bitcell is 2T1C DRAM (Fig.\ref{DRAMs}), that makes use of decoupled read and write ports, These are also called the gain cell DRAM, that make use of WWL and WBL to write a value onto SN and using RBL, RWL to read a value from SN. The write operation is similar to that of 1T1C. Prior to a read operation, the RBL is precharged to '0' in the case of PMOS being the read port transistor. During the read operation of a bitcell storing '0', RWL is turned ON and RBL is charged through the read-port transistor, which is further sensed using a sense amplifier. Similarly, in the case of storing '1', BL does not discharge as the PMOS read port transistor is OFF. However, the major disadvantage with this design is that the available margin for sensing the RBL voltage is limited during a read operation. This can be explained as: Assume there are 'm' rows in a single column that share the same RBL, with RWL corresponding to the unselected/selected rows in a column driven low/high and are storing a '0'/'1'. In such a scenario, RBL charges towards Vcc, because the selected row's WL is turned ON. However, as RBL is ramping up towards the threshold voltage of the read port transistor, the unselected rows experience a path towards gnd through the read port transistor, thereby discharging the RBL differential that was developed and constraining the sense margin to be equal to the threshold voltage of the read port transistor making it susceptible to process variations. In the case of NMOS being used as read port transistor, RBL is precharged to Vcc and the selected/unselected row's RWL driven high/low. In this case, the RBL voltage saturates at Vcc-Vt because of the leakage from the unselected rows, with Vt being the threshold voltage of the NMOS read port transistor \cite{DRAM_2T1C}. 
\par{\textbf{3T1C}}: Although 2T1C offers non-disruptive read mechanism, there was a need to improve the design in terms of resilience to process variations with increased sense margin. To accomplish this, 3T1C structure was proposed with 2 read port transistors (NMOS transistors) and 1 write port NMOS transistor, as seen in Fig.\ref{DRAMs}, similar to that of 8T SRAM. In this case, the write operation is similar to 1T1C design and the read operation is preceded by precharge operation, with RBL precharged to 'Vcc'. During a read operation, RWL is turned ON and RBL starts to discharge, with the additional transistor enabling read-out with higher sense margin by not allowing the unselected rows to discharge the RBL further, unlike 2T1C. This is accomplished by driving RWL of unselected rows to '0', ensuring that the RBL does not discharge through the unselected rows. However, both the 2T1C and 3T1C bitcells need separate refresh cycles, leading to increased power/energy, as the read operation does not imply a refresh operation because of the presence of a separate write and read bit lines (unlike the case of 1T1C) \cite{DRAM_3T1C}. 
\par The major issues with the DRAM are the long access latencies, reduced number of metal layers for routing the DRAM wires, thereby limiting the bandwidth of the memory array \cite{CDRAM_K}. To improve the performance of the DRAM design, embedded DRAMs that have the advantage of monolithic integration with the logic transistors were proposed, Furthermore, these have the advantage of stacking multiple layers and eDRAMs can be stacked in a 3D fashion, allowing increased bandwidth as compared to DRAMs, thereby enabling them to be used in last level caches. These can be designed in similar variants like 1T1C, 2T1C and 3T1C eDRAMs. However, the major disadvantage is that the eDRAMs use back end of the line (metal layers) based bitcell capacitor and scaling the capacitance to higher values is extremely difficult, thereby degrading the retention time of Silicon-based eDRAMs. Therefore, these offer the advantage of improved performance at the cost of lesser retention time. To further improve the retention time and reduce the leakage of eDRAMs, few novel materials based devices like Indium Gallium Zinc Oxide eDRAMs have been proposed, which have extreme low leakage with moderate ON currents have been proposed \cite{DRAM_1} \cite{DRAM_2}.
\begin{figure}[t]\centering
	\includegraphics[width=\textwidth]{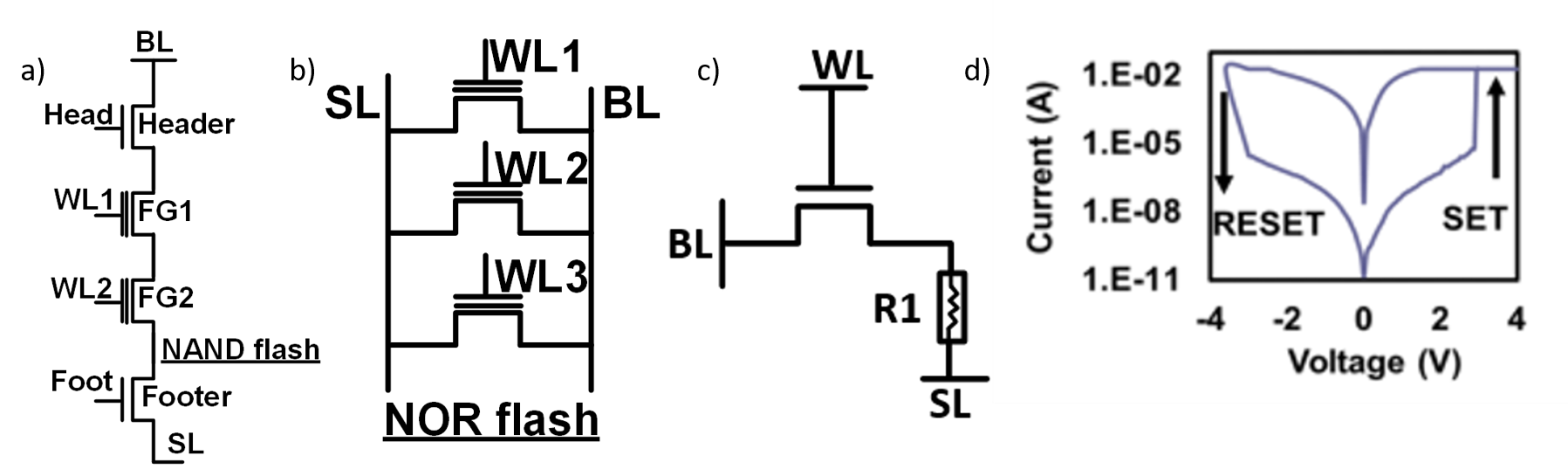}
	\caption{a) NAND and b) NOR flash, c) RRAM bitcell (1T1R) d) RRAM I-V characteristics \cite{RRAM_1} }\label{Flash}
\end{figure}
\subsection{Non-volatile memory technology}
Non-volatile memories are important for storing large amount of data that need to be saved for years together, as they retain data even in the absence of voltage supply. These are used extensively in hard disk/solid state drives(SSD) in the modern processors. The technology used in the SSDs is the NAND/NOR flash memory technology, as they have the advantage of easy integration, highly dense bitcell structure, stackable across 3D layers with minimal design effort, low cost/bit technology. This was a major replacement to old-age magnetic memory technology such as floppy, hard disk drives(HDD), which incur extremely long latency operations. The performance of NAND/NOR flash memories is better than the HDD, but worse than the already discussed SRAM/DRAM technology. This section begins by discussing the tradeoffs in the flash memory technologies and further extends to emerging non-volatile memory technologies including Resistive Random Access Memory, Magnetic Random Access Memory. 

\subsubsection{Flash memory}
\par{\textbf{NAND/NOR flash}}: NAND flash gets its name from the fact that the transistors are connected serially, similar to the pull-down transistors present in a CMOS NAND configuration. NAND has higher density, requires high write voltage, offers high write performance, low read performance as compared to NOR flash. The write operation is classified into programming (writing a '0') and erasing (writing a '1') by modulating the threshold voltage of the transistor that is being written into \cite{NAND_flash}. The NAND flash technology makes use of a floating and a separate control gate (CG), marked as 2 lines in Fig.\ref{Flash} transistor as the storage element, with CG trapping the electrons onto the floating gate using Fowler-Nordheim tunneling. In the case of programming, a high voltage is applied on the control gate and this allows the electrons in the channel between source to drain to get attracted onto the floating gate of the transistor, thereby depleting the channel of electrons, effectively increasing the threshold voltage of the transistor, resulting in a storage of '0' \cite{NAND_1}. In the case of storing a '1', the electrons that were trapped during the program phase are released by driving the control gate with a negative or 0 voltage. This results in the channel receiving more electrons, thereby decreasing the threshold voltage of the transistor. In the case of NAND flash, there are header and footer transistors that are used during the write and read operations \cite{NAND_flash_1}. \par During the write operation (program), the BL is driven to '0'(0V), WL corresponding to the selected row is driven high (for instance,$\sim$20V), with the other rows in the column driven with a voltage sufficient to pass the BL voltage onto the source of the row of transistors that are being written into and the gate to source voltage for selected cell is high to ensure that the tunneling happens. Furthermore, the header device is turned ON (for instance,$\sim$4V) to ensure that the BL voltage is passed onto the selected row. On the unselected columns, the BL is driven high (for instance, $\sim$4V), and as the NMOS transistors are not good at passing '1'(for instance,$\sim$4V), the header transistor goes into sub-threshold region, (as the drain = Vcc, source = Vcc-Vt, gate being at Vt). Thus, shutting down of the header transistor enables a non-disruptive operation on the unselected columns. In the case of programming, similar to other memory technologies, only a row is programmed, whereas in the case of erasing, a block of memory is erased. In other words, a memory array completely is erased \cite{NAND_flash_2}. 
\par During the erasing scenario, BL is initialized to 0 to ensure that all intermediate source and drain voltages are at 0 and are then left floating. SL can be left floating and the WL of the floating gate transistors are kept at 0 to ensure that the trapped electrons are ejected out of the floating gate into the channel for an entire memory array of cells. The header and footer word lines are turned ON, thereby ensuring that the source/drain voltages are 0 \cite{NAND_flash_book}. \par In the case of reading from NAND flash, the read is similar to programming in that the granularity of read is still a row of bitcells, unlike the case of erasing. The read of NAND flash can be accomplished by turning on the WLs of unselected rows to pass voltage, with BL driven high and SL driven low. The gate of the selected row is driven low and the current at the BL is measured across all columns to identify the read value from a row of cells. Furthermore, the major reason for NAND flash being used extensively for SSDs is the possibility of storing multiple levels in a bit-cell. There have been proposals suggesting storage of as high as, 1024 levels in a single bitcell. However, the major disadvantage of the flash technology is the voltages needed for operation can be extremely high in the order of 20V for programming, to capture electrons into the floating gate, limiting the power on the design and thereby it suffers from thermal bottleneck issues. The future of this technology lies in the ability to stack multiple layers with minimal coupling coefficient between layers. From a point of view of NOR flash, it is used only in microcontroller/internet of things based application space in older technology nodes and is still premature when it comes to usage in advanced technology, with a density as high as NAND flash \cite{NAND_2}.
\subsubsection{Resistive Random Access Memory(RRAM)}
    These are a class of emerging memory technology that rely on the principle of distinguishing the memory contents on the basis of resistance of the bitcell. The idea of using resistance based memory is helped by the fact that a variety of oxides exhibit resistive switching, i.e. the resistance changes as a function of the voltage applied between them (like HfO\textsubscript{2}, TiO\textsubscript{2}) and these binary metal oxides are easily compatible with CMOS\cite{RRAM_book}.  There are different RRAM variants like that of conductive bridge RRAM and oxide RRAM (detailed in this section). The conductive bridge RRAM (also known as electrochemical metallization memory) makes use of a metal ion for the formation of filament and rely on the movement of metal ions to determine the resistance of the device and subsequent switching. The oxide RRAM bitcell consists of metal (top electrode)-insulator-metal (bottom electrode) (R1 marked in Fig.\ref{Flash}), with the insulator being the above-mentioned oxides and rely on formation or breaking of oxide filament to store '1'(set) and '0'(reset), with Joule heating predicted to be the reason for filament rupture \cite{RRAM_future_2}. There are 2 types of RRAM namely the unipolar and bipolar RRAM, distinguished on the basis of voltages necessary to perform switching. It is important to understand the device characteristics to understand the applications and tradeoffs of the bitcell. In the case of bipolar switching, which is the most common usage of RRAM, the switching would involve applying positive voltages between the top and bottom electrode to perform "set operation" and applying negative voltages between the top and bottom electrode to perform the "reset operation". In the case of unipolar RRAM based switching mechanism, only positive voltage is sufficient to undergo set-reset and reset-set changes. Fig.\ref{Flash} describes the I-V characteristics for a bipolar switching RRAM, wherein the set voltage is high in the range of for instance,3-3.2V and the reset voltage is in the range of -ve for instance,3-3.2V\cite{RRAM_1}. The set and reset voltages need not be in the similar range and the set voltage can be considerably higher than the reset voltage, depending on the oxide insulator between the two electrodes. However, there are other devices of RRAM that have been proposed in the literature to reduce RRAM set/reset voltage to as low as for instance,1V\cite{RRAM_1}. With the initial state of the RRAM assumed to be in reset, the voltage is increased from 0 to the voltage necessary for setting, forming the filament necessary for conduction across RRAM, thereby allowing a low resistance path between the top and the bottom electrodes, shown as the set operation. Increasing the voltage beyond the set voltage strengthens the filament formation, and does not increase the current, marked by the saturation of current. In devices, wherein the current keeps increasing, a compliance current is maintained to restrict the increasing current to a certain threshold. In the case of decreasing the voltage from beyond set voltages towards 0, the current keeps decreasing, with a voltage of 0 still holding the filament, thereby having a non-zero current when the voltage comes back to 0. On decreasing the voltage in the negative directions, the current initially starts increasing till the voltage becomes equal to the reset voltage. On reaching the reset voltage, the current saturates and programs it into reset mode (that is, the filament formation is broken). Once the reset voltage is achieved, decreasing the voltage below the reset voltage breaks down the filament, thereby decreasing the current flowing through and increasing the resistance between the electrodes. These are non-volatile memory because the filament does not break/form if the current state is an already formed/broken filament, even when left without supplying voltage to retain the contents of the bitcell \cite{SRAM_2} \cite{RRAM_1}.
    \par It is important to understand the memory topology that is present to obtain a bitcell that can be used for storing. Similar to commodity DRAM structure, the RRAM based memory makes use of 1T1R structure wherein the access transistor has a WL that is responsible for accessing the bitcell. In the case of setting RRAM (storing '1'), a high voltage is applied on the BL node with SL node closer to 0, so that there is enough differential across the RRAM to obtain the filament, with WL turned ON. In the case of resetting RRAM (storing '0'), a high voltage is applied on the SL node with BL node closer to 0, breaking the filament formed by the set process. The design constraint in the case of setting is that the voltage at WL needs to be high enough to allow a high voltage through the access transistor, as NMOS based access transistors are not good at allowing '1' and saturates at Vcc-Vt. This should be taken into account when identifying the right BL voltage for setting. During the reset process, driving BL with a negative voltage is not preferred as the unselected rows in the same column would have WL equal to 0 and having a high negative voltage on the BL would mean a high negative voltage between the gate and the source of unselected rows, causing Gate Induced Drain Leakage (GIDL). Hence, SL is driven with a positive value, making sure that  the difference between voltage at the top and bottom node is negative, thereby resetting the filament formation. In the case of read, the voltage at RBL is driven to a predefined value and based on the amount of current flowing through the bitcell, the contents of the bitcell are identified. In the case of set mode, since the RRAM is in a lower resistance state, the amount of current sensed on the BL would be higher, indicating '1' and in the case of reset mode, since the RRAM is in a higher resistance state, the amount of current would be low, indicating '0'\cite{RRAM_future_3}. During a read operation, the constraint is that the read voltage should not be high enough to alter the state of the bitcell and just a "disturb voltage" is sufficient to accomplish a successful read. The ratio of high resistance state's resistance to low resistance state's resistance determines read margin of RRAM. In the case of RRAMs having high ratio of OFF to ON resistance, read margin is high and is resilient to process variations. Similar to the NAND flash, RRAM can be used effectively to store multiple levels in a single bitcell thereby adding to the advantage of dense bitcell. The disadvantages of RRAM involve the higher set/reset voltages, high read/write latency and lower endurance ($\sim$ 10\textsuperscript{6} cycles) as compared to the conventional SRAM/DRAM technology.
\begin{figure}[t]\centering
	\includegraphics[width=\textwidth]{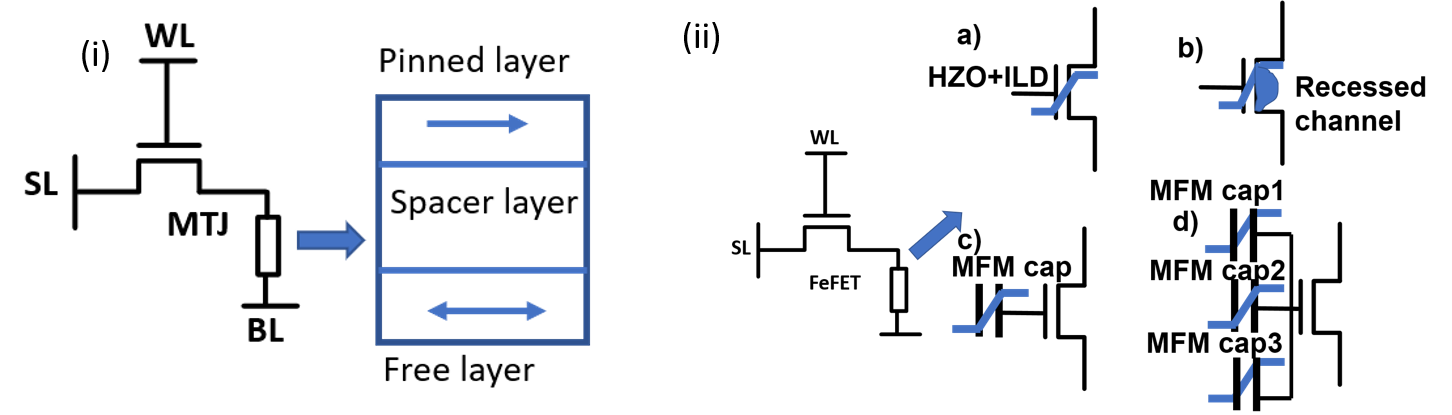}
	\caption{(i)STT-MRAM and (ii)FeFET bitcell} \label{MRAM_FeFET}
\end{figure}
\subsection{Magneto-resistive random access memory (MRAM)}
MRAM, like the RRAM also belongs to a class of memory technology that relies on the resistance of different states to store different contents onto the bitcell and builds concepts from magnetism/Spin Hall effect to modulate the bitcell resistance, hence the name magneto-resistive memory\cite{MRAM_future_1}. There are different types of MRAM namely the spin transfer torque (STT MRAM), Spin orbit torque MRAM (SOT MRAM) distinguished on the basis of the mechanism used for writing into the bitcell. The advantage of the bitcell as compared to RRAM is that the endurance of the bitcell is extremely high (in the order of $\sim$ 10\textsuperscript{15} cycles) with the write voltage slightly lower and lower write latency as compared to RRAM \cite{MRAM_future_2}. However, the disadvantage of this device lies in the complexity of integrating the different parts of the device together and the ratio of OFF to ON resistance is lower as compared to RRAM. However, the complexity of fabrication of the device has been taken care of, as STT MRAMs (Fig.\ref{MRAM_FeFET}) are ready for mass production. The magnetic tunnel junction (MTJ) which is the primary storage element for STT-MRAM, consists of 3 layers namely pinned, spacer layer and free layer with the relative orientation between the pinned and free layer determining the magneto-resistance of the device. The pinned layer has the magnetic moment pointed in one direction and does not change with application of external voltage. On the contrary, free layer's magnetic moment can be changed with the application of external voltage. If the magnetic moment of pinned layer and free layer point in the same direction, the magneto-resistance is low, and the resistance is high, when the magnetic moments point in the opposite direction. Similar to the case of RRAM, the direction of current determines the switching of MRAM and the current flow from pinned layer to free layer is responsible for switching the free layer from parallel (Low resistance - '1') to antiparallel (High resistance - '0') state. The current flow from free layer to pinned layer is responsible for switching from antiparallel to parallel state. A read operation is accomplished by applying a voltage at the BL and SL and the current through the MTJ (through single-ended sensing) is an indication of the magneto-resistance of the device \cite{MRAM_1} \cite{MRAM_2}.

\subsection{Ferroelectric Field Effect Transistor (FeFET)}
It belongs to a class of technology that makes use of capacitor to store data, similar to DRAM. It is a promising non-volatile memory (NVM) technology as it is dense, similar to RRAM, offers high speed as compared to RRAM and ease of manufacturing as they are compatible with mature CMOS technology nodes. The disadvantages of the existing design include the high program/erase voltage as compared to other NVM designs (order of 4V) and the lower retention time because of the innate depolarizing field in these devices\cite{FeFET_future_1}. FeFETs were formerly realized using a ferroelectric HZO (Hafnium Zinc Oxide) that is sandwiched between the metal and the typical oxide dielectric, as in the case of MOSFET, with the voltage division between the HZO capacitor and dielectric oxide (interlayer dielectric - ILD) for the voltage applied at the gate determining the bitcell content as shown in Fig.\ref{MRAM_FeFET}a). The major disadvantage is that the voltage drop across dielectric oxide is high and the voltage drop across HZO is minimal, thereby increasing the voltage needed for performing the write operation. Furthermore, there are innate fabrication difficulties to introduce the HZO layer between the gate and interlayer dielectric. To overcome the higher write voltage for FeFET, recessed FeFET was proposed, which increases the voltage drop across HZO, by making a geometry of the source-drain channel to be circular, concentrating the incoming electric field to a smaller region of area. FeMFET tries to overcome the integration difficulties faced by recessed FeFET, by integrating ferroelectric capacitor separately/independently at the gate (Fig.\ref{MRAM_FeFET}b)). This allows optimization of ferroelectric capacitor separately from that of MOSFET and write voltage can be reduced by decreasing the aspect ratio between ferroelectric capacitor and the interlayer dielectric of MOSFET Fig.\ref{MRAM_FeFET}c). However, the disadvantage of the design is that it introduces a floating node in between HZO capacitor and interlayer dielectric capacitance that is susceptible to noise, process variations and can reduce the retention time as the depolarizing field increases\cite{FeFET_future_2}.  Furthermore, it also affects the read voltage considerably, if there is leakage from the unselected cells in the same column. Furthermore, multiple ferroelectric capacitors can be connected in parallel at the gate to make sure that write voltage can be reduced because of the increase in capacitance of the ferroelectric capacitor, thereby making sure that most of the voltage drop is across the ferroelectric capacitor and not across the MOSFET. However, this approach is not scalable to larger voltage ranges and requires 3 cycles for a write operation. 
The write operation in these bitcells are accomplished by applying a voltage at the gate, and the voltage across the ferroelectric capacitor is an indication of the bitcell content. In the case of read, a read disturb voltage is applied on the top terminal of the ferroelectric capacitor, causing a voltage division, thereby enabling a higher voltage at the gate of the MOSFET, thereby implying a higher current through the FET which would imply '1' and '0' if the current through the MOSFET is lower \cite{FeFET_1} \cite{FeFET_2} \cite{FeFET_3}.
\section{Cryogenic memories}
Cryogenic computing has recently gained significant attention due to the emergence of superconducting processors, quantum computing control systems, and energy-efficient accelerators operating at temperatures ranging from 4\,K to 77\,K. However, conventional memory technologies exhibit substantially different characteristics at cryogenic temperatures compared to room-temperature operation. SRAMs generally demonstrate improved leakage characteristics and reduced sub-threshold conduction at low temperatures, enabling lower standby power consumption and improved data retention. Nevertheless, cryogenic operation also introduces challenges such as increased threshold voltage, degraded write margins, reduced read stability, and larger process variation sensitivity. Several works have proposed assist techniques, body-biasing methods, and emerging device integration strategies to improve SRAM robustness under cryogenic conditions \cite{ABI} \cite{UT_Thesis}. In particular, UTBB-SOI-based pseudo-static storage circuits have been proposed to exploit the ultra-low leakage characteristics of fully depleted SOI devices for cryogenic CMOS operation, enabling improved retention time and reduced standby energy under near-threshold voltage conditions. Similarly, DRAM and eDRAM technologies experience reduced leakage currents and potentially longer retention times at cryogenic temperatures, although capacitor behavior, sense-amplifier operation, and refresh management require redesign due to temperature-dependent variations in transistor and dielectric properties.

Emerging non-volatile memories have also shown strong potential for cryogenic systems because of their near-zero standby power and high density. RRAM devices exhibit improved retention and reduced leakage at low temperatures, although filament formation dynamics and switching voltages can vary significantly under cryogenic conditions. MRAM technologies, particularly spin-transfer torque MRAM (STT-MRAM), are attractive for cryogenic applications because magnetic tunnel junctions maintain non-volatility while offering high endurance and relatively fast switching behavior. Furthermore, FeFET-based memories have demonstrated promise due to their CMOS compatibility and low leakage operation, although ferroelectric polarization characteristics and switching behavior become highly temperature dependent. Beyond CMOS technologies, Josephson Junction Field-Effect Transistors (JJFETs) \cite{JJFET}\cite{DRAM} have also been explored for cryogenic computing applications because of their steep switching behavior and high noise margins at ultra-low temperatures. Such devices enable robust cryogenic logic operation while potentially reducing energy consumption for superconducting and quantum computing systems. These memory and logic technologies are increasingly being explored for cryogenic machine learning accelerators, superconducting computing platforms, and quantum computing control hardware, where energy efficiency, density, and low-temperature robustness are critical design requirements.

\section{Conclusion}
In this chapter, we discussed the different memory technologies starting from volatile memories like Static Random Access Memory (SRAM), Dynamic Random Access Memory (DRAM) to non-volatile memories like NAND/NOR flash, resistive random access memories (RRAM), magneto-resistive random access memories (MRAM), ferroelectric field effect transistor (FeFET) with specific reference to write, read and retention operations in each of these designs and the design constraints associated with them.


\end{document}